\documentclass{article}
\title{\bf Charged particles with spin in a gravitational wave and a uniform
magnetic field}
\author{Morteza Mohseni\thanks{email: m-mohseni@pnu.ac.ir}
\\{\small Physics Department, Payame Noor University, Tehran 19395-4697,
Iran}}
\begin{document}
\maketitle
\begin{abstract}
We study the motion of a pseudo-classical charged particle with
spin  in the space-time of a gravitational pp wave in the presence
of a uniform magnetic field.\\


\end{abstract}
\section{Introduction}

The theory of (pseudo)-classical charged spinning particles in
curved space-time in the presence of electromagnetic fields has
been a topic of interest in recent decades. The equations
governing the motion of these particles was obtained by dixon
\cite{dix1,dix2} and Souriau \cite{sou} by generalizing the
equations first obtained by Papapetrou \cite{pap} to describe
spinning test particles. These consist of a set of equations of
motion for the four-momentum and spin tensor of the particle, and
a third set of equations called the supplementary equations needed
to make the equations of motion complete. An extension of these
equations to the case where torsion fields are also included was
obtained in Refs. \cite{sol} and \cite{cog}. An alternative set of
equations was obtained by van Holten \cite{van}. The Dixon-Souriau
equations reduce to the van Holten equations whenever the
particle's four-momentum and four-velocity become co-linear. It
has also been shown that the Dixon-Souriau equations reduce to the
well known Bargmann-Michel-Telegdi equations in the limit of the
weak and homogeneous external field \cite{zer}.

The Dixon-Souriau equations was used in Ref. \cite{hoj} to study
the motion of charged spinning particles in a Kerr-Newman
background in the presence of electromagnetic fields. These
equations was also applied to the case of motion in a
Reissner-Nordstrom space-time in Ref. \cite{bin} (and also in Ref.
\cite{stu} with a different supplementary equation). A linearized
version of the Dixon-Souriau equations was used in Ref. \cite{pas}
to study charged spinning particles interacting with a
gravitational wave in the presence of a uniform magnetic field
(see also \cite{ver}). The equations of Ref. \cite{van} was used
in Ref. \cite{pra} to study the motion of a charged spinning
particle in the Schwarzschild space-time, in Ref. \cite{vir} to
study the motion in the Reissner-Nordstrom space-time, in Ref.
\cite{bal} to study the motion in the Taub-Nut space-time, and in
Ref. \cite{pravir} to study the motion in a uniform magnetic field
on a static space-time.

The aim of the present work is to study the motion of a charged
spinning test particles, in the space-time of a gravitational pp
wave in the presence of a uniform magnetic field. This is
motivated by the current interests in the dynamics of spinning
particles in curved space-times and its possible application to
the problem of the detection of gravitational waves \cite{nie2}.
We take the the electromagnetic field to be a uniform magnetic
field, both for calculational reasons and for its possible
astrophysical interests.

In our study of charged spinning particles, we use the equations
given in Ref. \cite{van}. The advantage of using these equations
instead of the Dixon-Souriau equations is their relative
simplicity compared with the latter ones. As we will see, we will
be able to find analytic solutions to the equations of motion
without resort to linear approximation. Since the external fields
considered here have very small amplitudes, we expect the results
to have the same physical content as the results of the
Dixon-Souriau equations.

\section{The Equations of Motion}

The equations describing the motion of a charged spinning particle
are as follows
\begin{eqnarray}
\frac{D\dot
x^\mu}{D\tau}&=&-\frac{1}{2m}{R^\mu}_{\nu\lambda\rho}S^{\lambda\rho}\dot
x^{\nu}+ \frac{q}{m}{F^\mu}_{\beta}\dot
x^{\beta}+\frac{q}{2m^2}S^{\kappa\rho}D^{\mu}F_{\kappa\rho}
,\label{e1}\\
\frac{DS^{\mu\nu}}{D\tau}&=&-\frac{q}{m}(S^{\mu\kappa}{F_\kappa}^\nu-S^{\nu\kappa}
{F_\kappa}^\mu),\label{e2}\\
\dot x_\mu S^{\mu\nu}&=&0,\label{e3}
\end{eqnarray}
where Greek indices run over the space-time dimensions, $\tau$ is
an affine parameter across the particle's word-line, $\dot
x^{\mu}$ is the 4-velocity of the particle, $S^{\mu\nu}$ are the
components of the spin tensor of the particle, $F^{\mu\nu}$ is the
electromagnetic tensor, $q$ and $m$ represent the charge and the
mass of the particle respectively, and
\begin{eqnarray*}
\frac{D\dot x^\mu}{D\tau}&=&\frac{d\dot x^{\mu}}{d\tau}+
\Gamma_{\lambda\nu}^{\mu}\dot x^{\lambda}\dot x^\nu,\\
\frac{DS^{\mu\nu}}{D\tau}&=&\frac{dS^{\mu\nu}}{d\tau}+\Gamma_{\lambda\rho}^{\mu}
\dot x^{\lambda}S^{\rho\nu}+\Gamma_{\lambda\rho}^{\nu}\dot x^{\lambda}S^{\mu\rho},\\
{R^\nu}_{\kappa\lambda\mu}&=&\partial_\lambda\Gamma^\nu_{\kappa\mu}
-\partial_\mu\Gamma^\nu_{\kappa\lambda}+\Gamma^\nu_{\lambda\rho}
\Gamma^\rho_{\kappa\mu}-\Gamma^\nu_{\mu\rho}\Gamma^\rho_{\kappa\lambda}.
\end{eqnarray*}
Equation (\ref{e3}) is the so called Pirani supplementary
condition. A more widely used supplementary condition is the so
called Tulczyjew supplementary condition which reads $$p_\mu
S^{\mu\nu}=0$$ in which $p^\mu$ is the particle's four-momentum.
For spinning particles, the four-momentum and the four-velocity
are not co-linear in general, but their difference is gauge
dependent and may be ignored in most practical cases \cite{nie2}.
By neglecting this difference the Tulczyjew condition reduces to
the Pirani condition and the Dixon-Souriau equations of motion
simplify to the equations (\ref{e1}) and (\ref{e2}). This
simplification enables us to find analytic expressions for the
trajectories of the particles and their spins without loss of
physical content.

The above equations give the components of $\dot x^\mu$ and
$S^{\mu\nu}$ and hence the trajectory of the particle. Also one
may convert the spin tensor $S^{\mu\nu}$ into the spin 4-vector
$S^\mu$ via the relation
\begin{equation}\label{e15}
S^\mu=\frac{1}{2\sqrt{-g}}\epsilon^{\mu\nu\kappa\lambda}\dot
x_{\nu} S_{\kappa\lambda}
\end{equation}
in which $\epsilon^{\mu\nu\kappa\lambda}$ is the alternating
symbol and we adopt $\epsilon^{1234}=+1.$ This sometimes gives a
better insight into the physics of the problem.

\section{The Space-Time}

The space-time we are interested in, is a plane gravitational wave
of general polarization and harmonic profile. We choose
$(u=t-z,v=t+z,x,y)$ as our coordinate system and represent them by
1,2,3,4 as indices respectively. In these coordinates the metric
is given by
\begin{equation}
ds^2=-dudv-K(u,x,y)du^2+dx^2+dy^2,\label{e5}
\end{equation}
where
\begin{equation}
K(u,x,y)=f_+(u)(x^2-y^2)+2f_\times(u) xy\label{e6a}
\end{equation}
and we take
\begin{eqnarray*}
f_+(u)&=&h\omega^2\sin(\omega u),\\
f_\times(u)&=& h\omega^2\cos(\omega u),
\end{eqnarray*}
$h$ being the dimensionless wave amplitude. We assume that a
constant magnetic field $B$ is present in this space-time and
neglect its effect on the space-time. We take the magnetic field
parallel to the direction of the wave propagation, that is ${\bf
B}=B{\hat k}$ and so in terms of $F^{\mu\nu}$, the only non
vanishing components are
\begin{equation}\label{e7}
{F^3}_4=-{F^4}_3=B.
\end{equation}

\section{The Trajectories and Spins}
We are interested in calculating the trajectory and spin evolution
of a charged spinning particle in the space-time given by the
metric (\ref{e5}) in the presence of a magnetic field given by
(\ref{e7}). We first set $\mu=1$ in equation (\ref{e1}) to obtain
\begin{eqnarray*}
\frac{d^2u(\tau)}{d\tau^2}=0
\end{eqnarray*}
where we choose the following solution
\begin{equation}\label{e10}
u=\tau
\end{equation}
that is we take $u$ to be the proper time along the particle's
world-line. Now by setting $\mu=1,\nu=3,4$ in equation (\ref{e2}),
we get
\begin{eqnarray*}
\frac{dS^{13}(\tau)}{d\tau}&=&\omega_0S^{14}(\tau)\\
\frac{dS^{14}(\tau)}{d\tau}&=&-\omega_0S^{13}(\tau).
\end{eqnarray*}
where $\omega_0=\frac{qB}{m}$ is the cyclotron frequency. The
solution to these equations reads
\begin{eqnarray}
S^{13}(\tau)&=&A\sin(\omega_0\tau+\theta)\label{e11},\\
S^{14}(\tau)&=&A\cos(\omega_0\tau+\theta)\label{e12}
\end{eqnarray}
where $A,\theta$ are constant. This means that these spin
components depend only on the magnetic field and their initial
values. Now by setting $\mu=3,4$ in equation (\ref{e1}) and using
equations (\ref{e11}) and (\ref{e12}) we obtain
\begin{eqnarray*}
\frac{d^2x(\tau)}{d\tau^2}=&-&h\omega^2\sin(\omega\tau)x(\tau)-h\omega^2\cos(\omega\tau)y(\tau)
\nonumber\\&+&\omega_0\frac{dy(\tau)}{d\tau}+\frac{Ah\omega^2}{m}\cos((\omega_0-\omega)\tau
+\theta),\\
\frac{d^2y(\tau)}{d\tau^2}=&&h\omega^2\sin(\omega\tau)y(\tau)-h\omega^2\cos(\omega\tau)x(\tau)
\nonumber\\&-&\omega_0\frac{dx(\tau)}{d\tau}+\frac{Ah\omega^2}{m}\sin((\omega_0-\omega)\tau
+\theta).
\end{eqnarray*}
These can be solved for $\chi(\tau)=x(\tau)+iy(\tau)$ resulting in
\begin{equation}\label{e31a}
\chi(\tau)=e^{-\frac{i\omega\tau}{2}}\left(Pe^{\alpha\omega\tau}+Qe^{-\alpha\omega\tau}\right)
+i\frac{A}{m}e^{i\theta}e^{-i\delta\omega\tau}
\end{equation}
which is valid for $|\delta-\frac{1}{2}|<2h$, and
\begin{eqnarray}\label{e31b}
\chi(\tau)=&&Pe^{i\alpha_1\omega\tau}+Qe^{i\alpha_2\omega\tau}+i\frac{A}{m}
e^{i\theta}e^{-i\delta\omega\tau}\nonumber\\&+&i\Delta_1{\overline
P}e^{-i(1+\alpha_1)\omega\tau}+i\Delta_2{\overline
Q}e^{-i(1+\alpha_2)\omega\tau}
\end{eqnarray}
valid for other values of $\delta$ (we assume that
$h<\frac{1}{4}$). Here, $P=P_1+iP_2,Q=Q_1+iQ_2$ are constants to
be determined from $\chi(0),{\dot\chi}(0)$, and
\begin{eqnarray*}
\delta&=&\frac{\omega_0}{\omega},\\
\alpha&=&\frac{1}{2}\sqrt{2\delta-2\delta^2-1+2\sqrt{\delta^4-2\delta^3+\delta^2+4h^2}},\\
\alpha_1&=&\frac{-1+\sqrt{2\delta^2-2\delta+1+2\sqrt{\delta^4-2\delta^3+\delta^2+4h^2}}}{2},\\
\alpha_2&=&\frac{-1+\sqrt{2\delta^2-2\delta+1-2\sqrt{\delta^4-2\delta^3+\delta^2+4h^2}}}{2},\\
\Delta_1&=&\frac{h}{(1+\alpha_1)^2-(1+\alpha_1)\delta},\\
\Delta_2&=&\frac{h}{(1+\alpha_2)^2-(1+\alpha_2)\delta}.
\end{eqnarray*}
The equation for $v(\tau)$ may be obtained either by setting
$\mu=2$ in equation (\ref{e1}) and solving the resulting equation
for $v$, or more directly by integrating the following relation
$${\dot v}(\tau)=1-K(u,x,y)+({\dot x}(\tau))^2+({\dot y}(\tau))^2$$
which in turn is a consequence of the relation $\dot x_\mu\dot
x^\mu=-1.$ Thus we have
\begin{equation}\label{e14}
v(\tau)=\tau+\int\left({\dot x}^2(\tau)+{\dot
y}^2(\tau)+K(u,x,y)\right)d\tau.
\end{equation}
Let us now return to the equation (\ref{e2}), where for
$\mu=3,\nu=4$ we have
$$\frac{dS^{34}(\tau)}{d\tau}=Ah\omega^2(x(\tau)\sin((\delta-1)\omega\tau+\theta)
-y(\tau)\cos((\delta-1)\omega\tau)+\theta).$$ Now by substituting
$x(\tau)$ and $y(\tau)$ into this equation, one may solve it for
$S^{34}(\tau)$ to obtain
\begin{eqnarray}\label{e41a}
S^{34}(\tau)=&&\frac{-Ah\omega}{\alpha^2+(\delta-\frac{1}{2})^2}\left((P_1(\delta-\frac{1}{2})
+P_2\alpha)e^{\alpha\omega\tau}\cos((\delta-\frac{1}{2})\omega\tau+\theta)+\nonumber\right.\\&&
\hspace{28mm}(Q_1(\delta-\frac{1}{2})-Q_2\alpha)e^{-\alpha\omega\tau}\cos((\delta-\frac{1}{2}
)\omega\tau+\theta)+\nonumber\\&&\hspace{28mm}(P_2(\delta-\frac{1}{2})-P_1\alpha)e^{\alpha\omega\tau}
\sin((\delta-\frac{1}{2})\omega\tau+\theta)+\nonumber\\&&\hspace{28mm}\left.(Q_2(\delta-\frac{1}
{2})+Q_1\alpha)e^{-\alpha\omega\tau}\sin((\delta-\frac{1}{2})\omega\tau+\theta)\right)\nonumber
\\&&-\frac{A^2h\omega}{m(1-2\delta)}\sin((1-2\delta)\omega\tau)+C
\end{eqnarray}
valid for $|\delta-\frac{1}{2}|<2h$. For other values of $\delta$
we have
\begin{eqnarray}
S^{34}(\tau)=&&Ah\omega\left(\frac{P_1}{\alpha_1+1-\delta}\cos((\alpha_1+1-\delta)\omega\tau
-\theta)-\nonumber\right.\\&&\hspace{12mm}\frac{P_2}{\alpha_1+1-\delta}\sin((\alpha_1+1-\delta)
\omega\tau-\theta)-\nonumber\\&&\hspace{12mm}\frac{P_2\Delta_1}{\alpha_1+\delta}\cos((\alpha_1
+\delta)\omega\tau+\theta)-\nonumber\\&&\hspace{12mm}\frac{P_1\Delta_1}{\alpha_1+\delta}
\sin((\alpha_1+\delta)\omega\tau+\theta)+\nonumber\\&&\hspace{12mm}\frac{Q_1}{\alpha_2+1-\delta}
\cos((\alpha_2+1-\delta)\omega\tau-\theta)-\nonumber\\&&\hspace{12mm}\frac{Q_2}{\alpha_2+1-
\delta}\sin((\alpha_2+1-\delta)\omega\tau-\theta)-\nonumber\\&&\hspace{12mm}\frac{Q_2
\Delta_2}{\alpha_2+\delta}\cos((\alpha_2+\delta)\omega\tau+\theta)-\nonumber\\&&\hspace{12mm}
\left.\frac{Q_1\Delta_2}{\alpha_2+\delta}\sin((\alpha_2+\delta)\omega\tau+\theta)\right)
\nonumber\\&-&\frac{A^2h\omega}{m(1-2\delta)}\sin((1-2\delta)\omega\tau)+C\label{e41b}
\end{eqnarray}
in which C is a constant depending on the initial value of
$S^{34}(\tau)$. The other components of $S^{\mu\nu}$ may be
obtained from equation (\ref{e3}). They are as follows
\begin{eqnarray}
S^{12}(\tau)&=&2{\dot x}(\tau)S^{13}(\tau)+2{\dot
y}(\tau)S^{14}(\tau),\label{e16}\\
S^{23}(\tau)&=&-2{\dot y}(\tau)S^{34}(\tau)-\left({\dot
v}(\tau)+2K(u,x,y)\right)S^{13}(\tau),\label{e17}\\
S^{24}(\tau)&=&2{\dot x}(\tau)S^{34}(\tau)-\left({\dot
v}(\tau)+2K(u,x,y)\right)S^{14}(\tau).\label{e18}
\end{eqnarray}
Finally, making use of the relation (\ref{e15}), we obtain
\begin{eqnarray}
S^1(\tau)&=&-S^{34}(\tau)+{\dot x}(\tau)S^{14}(\tau)
-{\dot y}(\tau)S^{13}(\tau),\\
S^2(\tau)&=&2S^{34}(\tau)+{\dot v}(\tau)S^1(\tau),\\
S^3(\tau)&=&S^{14}(\tau)+{\dot x}(\tau)S^1(\tau),\\
S^4(\tau)&=&-S^{13}(\tau)+{\dot y}(\tau)S^1(\tau).
\end{eqnarray}
The explicit form of these components is presented in the
appendices for some values of $\delta$.

\section{Discussion}

If we set $B=0$, that is if the magnetic field is absent, we have
a spinning particle moving in a gravitational wave background.
This has been studied extensively in Refs. \cite{nie2}-\cite{moh}.
For the case of the vanishing magnetic field, the equations
obtained here reduce to those obtained in Ref. \cite{moh}.

An interesting feature of motion of a charged spinning particle in
the external fields considered here is that spin-electromagnetic
interaction does not affect the particle's trajectory directly,
this can be seen from the absence of the last term of equation
(\ref{e1}) in the subsequent equations for $\chi(\tau)$. However
in the case that the spin has an initial transverse component,
$A\neq 0$, this interaction affects the spin and this in turn
affects the trajectory. If the particle has no initial transverse
spin components, its spins remains unaffected. This can be seen by
letting $A=0$ in equations (\ref{e11}), (\ref{e12}), and
(\ref{e41a}),(\ref{e41b}). Also in this case, if the particle is
sitting initially at the origin of coordinates, it remains in this
coordinate point at other times. Another interesting feature of
these solutions is that they consist only of oscillatory terms,
except for the interval $\frac{1}{2}-2h<\delta<\frac{1}{2}+2h$
which is narrow for small $h$. For the values of $\delta$ inside
this interval, the oscillatory terms are modulated by slowly
varying exponential terms $e^{\pm h\omega\tau}$ . For $\delta=1$
the coefficients $\alpha_1,\alpha_2$ become the same as those
obtained in Ref. \cite{moh}, see the appendices.

For an object like an electron (treated classically) moving in a
uniform magnetic field of the same typical magnitude as that of
the earth, and for the gravitational wave frequencies ranging from
$1 Hz$ up to $10^4 Hz$, the parameter $\delta$ ranges from an
order of magnitude of $10^7$ down to $10^{3}$, and so the effect
of the gravitational wave is suppressed by the magnetic field. For
gravitational waves with frequencies lower than $1 Hz$, $\delta$
exceeds $10^7$. For weaker magnetic fields such as cosmological
magnetic fields, which could be as weak as $10^{-12} T$, this
parameter, again for an electron, is of an order of magnitude of
$10\sim 10^{-4}.$ Thus $\delta=1$ corresponds to situations such
as the case where an electron moves in a weak magnetic field (of
say, $10^{-9}T$) and a gravitational wave of frequency of say
$10^2 Hz$. The calculation of relative acceleration of charged
spinning particles in gravitational waves and  magnetic fields
will be reported elsewhere.

\section*{Acknowledgments}

The author would like to thank the Abdus Salam ICTP (Trieste)
where part of this work was done.

\appendix
%\section{Appendices}
\section{The Case of $\delta=\frac{1}{2}$}
For $\delta=\frac{1}{2}$ we have
$$\alpha=\frac{1}{2}\sqrt{-\frac{1}{2}+\frac{1}{2}\sqrt{1+64h^2}}\approx 2h$$
and so neglecting $h^2$ and higher order terms, we obtain
\begin{eqnarray}
x(\tau)=&&\left(P_1e^{2h\omega\tau}+Q_1e^{-2h\omega\tau}\right)
\cos\left(\frac{\omega\tau}{2}\right)\nonumber+\\&&\left(P_2e^{2h\omega\tau}
+Q_2e^{-2h\omega\tau}\right)\sin\left(\frac{\omega\tau}{2}\right)+
\frac{A}{m}\sin\left(\frac{\omega\tau}{2}-\theta\right),\label{e51a}
\\y(\tau)=&&\left(P_2e^{2h\omega\tau}+Q_2e^{-2h\omega\tau}\right)
\cos\left(\frac{\omega\tau}{2}\right)\nonumber-\\&&\left(P_1e^{2h\omega\tau}
+Q_1e^{-2h\omega\tau}\right)\sin\left(\frac{\omega\tau}{2}\right)+
\frac{A}{m}\cos\left(\frac{\omega\tau}{2}-\theta\right),\label{e51b}\\
S^{34}(\tau)=&&\frac{1}{2}A\omega\left((P_1\sin\theta-P_2\cos\theta)
e^{2h\omega\tau}\right.\nonumber+\\&&\left.(Q_2\cos\theta-Q_1\sin\theta)e^{-2h\omega\tau}
\right)-\frac{A^2h\omega^2}{m}\tau+C.
\end{eqnarray}
If we choose $\chi(0)={\dot\chi}(0)=0$, we obtain
\begin{eqnarray}
x(\tau)&=&\frac{A}{m}\sin\left(\frac{1}{2}\omega\tau-\theta\right)(1-\cosh(2h\omega\tau))
,\label{e251}\\
x(\tau)&=&\frac{A}{m}\sin\left(\frac{1}{2}\omega\tau-\theta\right)(1-\cosh(2h\omega\tau))
,\label{e252}\\
S^{34}(\tau)&=&\frac{A^2\omega}{2m}\left(\sinh(2h\omega\tau)-2h\omega\tau\right)+C,\label{e253}
\end{eqnarray}
and hence
\begin{eqnarray}
S^1(\tau)&=&-\frac{A^2\omega}{2m}\left(\sinh(2h\omega\tau)-2h\omega\tau\right)-C
,\label{e255a}\\
S^2(\tau)&=&\frac{A^2\omega}{2m}\left(\sinh(2h\omega\tau)-2h\omega\tau\right)+C
,\label{e255}\\
S^3(\tau)&=&A\cos(\omega\tau+\theta),\label{e256}\\
S^4(\tau)&=&-A\sin(\omega\tau+\theta).\label{e257}
\end{eqnarray}
We also have
\begin{eqnarray}
S^z(\tau)=\frac{1}{2}(S^2(\tau)-S^1(\tau))
\approx\frac{A^2\omega}{2m}\left(\sinh(2h\omega\tau)-2h\omega\tau\right)+C.
\end{eqnarray}

\section{The Case of $\delta=1$}

For $\delta=1$ we have
$$\alpha_1=\frac{-1+\sqrt{1+4h}}{2},\hspace{3mm}\alpha_2=\frac{-1+\sqrt{1-4h}}{2}
,\hspace{3mm}\Delta_1=-\Delta_2=1$$ and hence
\begin{eqnarray}\label{e52}
\chi(\tau)=&&Pe^{i\alpha_1\omega\tau}+i{\overline
P}e^{-i(1+\alpha_1)\omega\tau}+Qe^{i\alpha_2\omega\tau}\nonumber\\&-&i{\overline
Q}e^{-i(1+\alpha_2)\omega\tau}+i\frac{A}{m}e^{-i(\omega\tau-\theta)}
\end{eqnarray}
In addition, if we let $\chi(0)={\dot\chi}(0)=0$, we have
\begin{eqnarray}\label{e53}
\chi(\tau)=&&\frac{A}{2m}\left(-\frac{(1+\alpha_1)e^{-i\theta}+i\alpha_1e^{i\theta}}
{1+2\alpha_1}e^{i\alpha_1\omega\tau}\right.\nonumber\\&&\hspace{10mm}-\frac{\alpha_1e^{-i\theta}
+i(1+\alpha_1)e^{i\theta}}{1+2\alpha_1}e^{-i(1+\alpha_1)\omega\tau}\nonumber\\&&\hspace{10mm}+
\frac{(1+\alpha_2)e^{-i\theta}-i\alpha_2e^{i\theta}}{1+2\alpha_2}e^{i\alpha_2\omega\tau}
\nonumber\\&&\hspace{10mm}+\frac{\alpha_2e^{-i\theta}-i(1+\alpha_2)e^{i\theta}}{1+2\alpha_2}
e^{-i(1+\alpha_2)\omega\tau}\nonumber\\&&\hspace{10mm}\left.+2ie^{-i(\omega\tau-\theta)}\right).
\end{eqnarray}
For small $h$ we have
\begin{eqnarray}
x(\tau)&=&\frac{Ah}{m}(\cos\theta-\cos(\omega\tau+\theta)-\frac{A}{m}\sin\theta\sin(h\omega\tau)
,\label{e54}\\
y(\tau)&=&\frac{Ah}{m}(-\sin\theta+\sin(\omega\tau+\theta))-\frac{A}{m}\cos\theta
\sin(h\omega\tau),\label{e55}\\
S^{34}(\tau)&=&C\label{e65},
\end{eqnarray}
and hence
\begin{eqnarray}
S^1(\tau)=&-&C+\frac{A^2h\omega}{m}\sin\theta\left(\sin(\omega\tau+\theta)-\cos(\omega\tau
+\theta)\right),\label{e57}\\
S^2(\tau)=&&C+\frac{A^2h\omega}{m}\sin\theta\left(\sin(\omega\tau+\theta)-\cos(\omega\tau
+\theta)\right)\label{e58}\\
S^3(\tau)=&&A\cos(\omega\tau+\theta)-\frac{hCA\omega}{m}\left(\sin(\omega\tau+\theta)-
\sin\theta\right),\label{e59}\\
S^4(\tau)=&-&A\sin(\omega\tau+\theta)-\frac{hCA\omega}{m}\left(\cos(\omega\tau+\theta)-
\cos\theta\right).\label{e60}
\end{eqnarray}
We also obtain
\begin{equation}
S^z(\tau)=\frac{1}{2}\left(S^2(\tau)-S^1(\tau)\right)=C.
\end{equation}
which can be used to determine constant $C$ in terms of the spin
component along the direction in which the wave propagates.

\end{document}